\documentclass[floatfix,pra,twocolumn,showpacs,amsmath,amssymb,letterpaper,groupaddresses,superscriptaddress]{revtex4}
\usepackage{times}
\usepackage{latexsym}
\usepackage{graphicx}
\usepackage{verbatim,times,bbm}

\begin{document}

\title{Removing correlations in signals transmitted over a quantum memory channel}

\author{Cosmo Lupo}
\affiliation{School of Science and Technology, University of Camerino, I-62032 Camerino, Italy}

\author{Laleh Memarzadeh}
\affiliation{Physics department, Sharif University of Technology, P.O. Box 11155-9161, Tehran, Iran}

\author{Stefano Mancini}
\affiliation{School of Science and Technology, University of Camerino, I-62032 Camerino, Italy}
\affiliation{INFN-Sezione di Perugia, I-06123 Perugia, Italy}

\begin{abstract}
We consider a model of bosonic memory channel, which induces correlations among the transmitted signals.
The application of suitable unitary transformations at encoding and decoding stages allows
the complete removal of correlations, mapping the memory channel into a memoryless one.
However, such transformations, being global over an arbitrary large number of bosonic modes,
are not realistically implementable.
We then introduce a family of efficiently realizable transformations which can be used to
partially remove correlations among errors, and we quantify the reduction of the gap with memoryless channels.
\end{abstract}


\maketitle

\section{Introduction}

Any physical medium that can store or transfer quantum degrees of freedom
can be formally described as a quantum channel, that is, a completely positive
and trace-preserving map on the set of quantum states.
In information theory, quantum channels are mainly intended as means to
convey either classical or quantum information \cite{QInf}.
They model the noisy interaction with the environment, causing losses, dephasing
and decoherence.
The action of the noise is usually to degrade the information carried by the 
physical system, hence reducing the performance of quantum communication protocols.
A standard strategy to protect the information against noise is the use of quantum 
error correcting codes \cite{gotty1}.
These latter exploit redundancy and dictate the way information should be encoded/decoded
in quantum systems.
In this scenario, a common assumption is that the noise acts identically and 
independently at each use of the quantum channel.
That is, when a train of signals is sent through the quantum channel,
each signal independently experiences the noisy transformation.
A quantum channel with this property is said to be memoryless.

Recently, attention has been devoted to quantum memory channels as well
(see e.g. \cite{KW2} for discrete channels and \cite{CVmem} for continuous channels), where
the action of the noise at different channel uses is either non indentical or non
independent.
From a physical point of view, that is the case when the typical environmental
relaxation times are comparable with the time delay between two signals.
From an information theoretical point of view, the presence of correlations
in the noise can substantially reduce the efficacy of standard quantum error
correcting codes \cite{qcodeneg}.

Then there is a twofold way to deal with information transmission over memory channels:
either consider new encoding/decoding procedures accordingly to codes specifically devised 
for correlated errors \cite{qcodenew}, or try to remove correlations and then apply standard 
encoding/decoding procedures.
Here, we shall investigate the second line which results unexplored so far. 
We shall restrict our attention to bosonic memory channels.

In Ref.\cite{trans} it has been discussed a technique, called \emph{memory unravelling}, 
that allows the removal of correlations in a large class of bosonic memory channels. 
Actually the channels can be traced back to memoryless ones by global unitary transformations 
prepended and appended to the channel map.
However this is just a mathematical trick useful for evaluating capacites. 
For practical transmission rates it is unreasonable to realize transformations involving a large 
number of modes (channel uses).
We shall hence introduce unitaries that can be implemented efficiently but only allow a partial 
reduction of the correlations in the memory channel.

Specifically we shall consider the model of lossy bosonic memory channel introduced and
characterized in \cite{LGM} for which memory unravelling applies.
It is known \cite{vit} that for the corresponding memoryless attenuating channel
a standard encoding/decoding to get close to the classical capacity is given by (single mode) 
coherent states and dyne measurements  \cite{dyne}.
Then, we shall evaluate the performance of such encoding/decoding
procedure upon efficient, but partial, removal of correlations.
In particular we quantify the reduction of
the gap with memoryless channels by means of input-output mutual information.

The article will proceed as follows. In Sec.\ \ref{model} we shall introduce the model of
bosonic memory channel under scrutiny; in Sec.\ \ref{Remove} we shall consider the
unitary pre-processing and post-processing transformations allowing the complete removal of
correlations; in Sec.\ \ref{partial} we shall introduce an efficient scheme for partially
reducing the correlations based on unitaries acting on a finite number of bosonic modes.
Numerical results about the reduction of the gap with memoryless channels in the case of
chaining unitaries acting on two and three modes will be presented in Sec.\ \ref{Res}.
Conclusions will be drawn in Sec.\ \ref{end}.

\section{The memory channel}\label{model}

As a case study, we consider the Gaussian memory channel
introduced and characterized in \cite{LGM}, which models memory
effects in linear attenuating media, acting on a set of bosonic
modes (e.g., normal modes of the electromagnetic field).
In the same spirit of the Refs.~\cite{KW2,bowen}, the memory channel
under scrutiny is defined as a sequence of elementary transformations,
which are concatenated through a {\em memory} mode.
An elementary transformation involves a pair of ingoing bosonic modes,
described through the ladder operators $\{ a, a^\dag \}$ and $\{ c, c^\dag \}$,
and the corresponding outgoing modes $\{ b, b^\dag \}$, $\{ d, d^\dag\}$.
The modes are coupled via a unitary interaction described by an
operator of the form
$U_\mathrm{BS}=\exp{(\theta a^\dag c - \mathrm{h.c.})}$, modeling an
exchange interaction mixing the two modes.
In the following we assume the parameter $\theta$ to be real and positive,
yielding the following Heisenberg-picture transformations on the ingoing modes
\begin{subequations}\label{BSX}
\begin{eqnarray}
b = {U_\mathrm{BS}}^{\hspace{-0.05cm}\dag} \, a \, U_\mathrm{BS} = \cos{\theta} \, a -\sin{\theta} \, c \, , \\
d = {U_\mathrm{BS}}^{\hspace{-0.05cm}\dag} \, c \, U_\mathrm{BS} =
\cos{\theta} \, c +\sin{\theta} \, a \, ,
\end{eqnarray}
\end{subequations}
together with their hermitian conjugates.
The quantity $\cos^2{\theta}$ is usually referred to as the
transmissivity parameter.

To define the action of the quantum memory channel, we consider a
sequence of $n$ consecutive channel uses. Such a sequence is
associated with a collection of $n$ bosonic modes with ladder
operators $\{ a_j, {a_j}^{\hspace{-0.05cm}\dag} \}_{j=1,\dots n}$,
representing the channel inputs, and the corresponding channel
outputs associated with the ladder operators $\{ b_j,
{b_j}^{\hspace{-0.05cm}\dag} \}_{j=1,\dots n}$. We model the channel
environment as a collection of environmental modes with $\{ e_j,
{e_j}^{\hspace{-0.05cm}\dag} \}_{j=1,\dots n}$. The {\it memory} modes,
which account for the flux of information from one channel use to the
following, are associated to the ladder operators $\{ m_j,
{m_j}^{\hspace{-0.05cm}\dag} \}_{j=1,\dots n}$.

\begin{figure}[htb]
\centering
\includegraphics[width=0.4\textwidth]{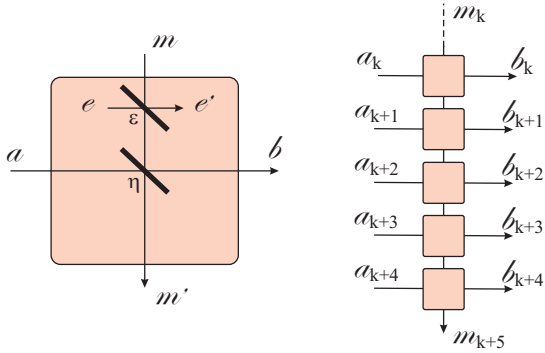}
\caption{Left: a single use of the attenuating
memory channel is described as an elementary transformation which is
the concatenation of two beam-splitters, respectively characterized
by the transmissivities $\epsilon$ and $\eta$. The first beam-splitter
couples the memory mode with the local environment, the second one
mixes the memory mode with the input mode. Right: $n$ uses of the
memory channel are described as the concatenation of the
elementary transformation \cite{KW2,bowen}. The concatenation is
obtained by identifying, for any $k$, the outgoing memory mode at
the $k$-th channel use with ingoing memory mode at the $(k+1)$-th.
The memoryless limit is obtained by cutting the information flow
through the memory mode, i.e., setting $\epsilon=0$.}
\label{lossym}
\end{figure}

A single use of the memory channel is modeled as the composition of
two beam-splitters with transmissivities $\epsilon$, $\eta$
(transmissivities are within the interval $[0,1]$), as depicted in Fig.\ \ref{lossym}.
In the Heisenberg picture, it transforms the ladder operators according to
\begin{subequations}\label{lossy1}
\begin{align}
m' & = \sqrt{\epsilon\eta} \, m + \sqrt{1-\eta} \, a + \sqrt{(1-\epsilon)\eta} \, e \, , \label{lossy1m}\\
b & = \sqrt{\eta} \, a - \sqrt{(1-\epsilon)(1-\eta)} \, e \, - \sqrt{\epsilon(1-\eta)} \, m \, ,
\end{align}
\end{subequations}
together with the hermitian conjugate relations.

The action of the memory channel upon $n$ uses is obtained by
identifying the outgoing memory mode at each channel use with the
ingoing memory mode at the following one (see Fig.\ \ref{lossym}).
In the Heisenberg picture, by iteration of Eq.s~(\ref{lossy1}), we have
\begin{align}
b_k = & \sqrt{\eta} \, a_k - \sqrt{\epsilon}(1-\eta) \sum_{j=1}^{k-1} \left(\sqrt{\epsilon\eta}\right)^{k-j-1} a_{j} \nonumber\\
& -\sqrt{(1-\epsilon)(1-\eta)} \, \sum_{j=1}^k \left(\sqrt{\epsilon\eta}\right)^{k-j} e_{j} \nonumber \\
& + \sqrt{\epsilon(1-\eta)}\left(\sqrt{\epsilon\eta}\right)^{k-1} m_1 \, , \label{bk}
\end{align}
for $k=1, \dots n$.
From Eq.\ (\ref{bk}) it follows that, for each $k$, the output mode
operator $b_k$ is a function of the corresponding input mode
operator $a_k$ and of the input mode operators in its past: $a_j$,
with $k > j$. Hence, a certain amount of information from the $j$-th
input flows to the $k$-th output, for $k > j$. That is, the input
signals from different channel uses interfere at the channel outputs,
leading to memory effects in the quantum channel.

Equation (\ref{bk}) can be concisely rewritten in the following form:
\begin{equation}\label{linear_1}
b_k = \sum_j f_{kj} a_j + \sum_j g_{kj} e_j + t_k m_1 \, .
\end{equation}

Finally, to exhaustively define the channel model, we have to fix
the initial state of the local environmental modes $\{ e_j, e_j^\dag
\}$. Different choices for the environment states lead to channels
with different features. In the following we assume the local
environments to be in the vacuum state.
With this choice for the environmental states the quantum and classical
capacities of the memory channel can be computed exactly \cite{LGM}.

The transmissivity
can be related to the ratio between the time delay
$\Delta t$ between to successive channel uses and the typical
relaxation time $\tau_{rel}$ of the channel environment~\cite{VJP}: for
instance we may identify $\epsilon \simeq \exp{(-\Delta t/\tau_{rel})}$.
In particular, the model reduces to a memoryless attenuating
channel~\cite{vit} for $\epsilon=0$ (the input $a_j$
only influences the output $b_j$), and to a channel with perfect
memory \cite{bowen} for $\epsilon=1$ (all $a_j$'s interacts {\em only}
with the memory mode). These two limiting settings respectively
correspond to the regime $\Delta t \gg \tau_{rel}$, and $\Delta t \ll
\tau_{rel}$. Intermediate configurations are associated with values $\epsilon
\in (0,1)$ and correspond to {\it inter-symbol interference}
channels, for which the previous input states affect the action of
the channel on the current input~\cite{BDM}.


\section{Removing correlations}\label{Remove}

As it has been shown in \cite{LGM}, a unitary transformation exists
which completely remove the correlations in the memory channel.
Such optimal transformation can be constructed by considering the coefficients
$f_{kj}$ appearing in Eq.\ (\ref{linear_1}), which express the linear relations
between the input and output field operators.
For any $n$, these coefficients define a $n \times n$ matrix $f$, which admits a singular
value decomposition:
\begin{equation}
f = \mathcal{V}^\dag \, \sqrt{\Delta} \, \mathcal{U} \, ,
\end{equation}
where $\mathcal{U}$, $\mathcal{V}$ are $n \times n$ unitary matrices and
$\Delta=\mathrm{diag}(\eta_1, \eta_2, \dots \eta_n)$ is diagonal with non-negative entries.
The unitary matrices play the role of unraveling the correlations.
In the Heisenberg picture, it maps the input and output operators in
\begin{align}
A_k & := \sum_j {\mathcal{U}}_{kj} \, a_j \, , \label{input} \\
B_k & := \sum_j {\mathcal{V}}_{kj} \, b_j \, . \label{output}
\end{align}
It follows that these collective input and output field operators satisfy the identities
\begin{equation}\label{linear_2}
B_k = \sqrt{\eta_k} A_k + \sum_{l,j} {\mathcal{V}}_{kl} g_{lj} e_j + \sum_{l} {\mathcal{V}}_{kl} t_l m_1 \, ,
\end{equation}
from which it is apparent that the $k$-th collective output signal is only
influnced by the $k$-th input signal, i.e., the correlations due to the
inter-symbol interference have been completely removed.
Another source of correlated noise can be represented by the noisy terms
associated to the environmental modes with operators $e_j$, $m_1$.
However, if these modes are not populated (or, if they are in a thermal state)
their presence does not induce additional correlation terms.
In fact, Eq.\ (\ref{linear_2}) can be written in the following form \cite{LGM}:
\begin{equation}\label{linear_3}
B_k = \sqrt{\eta_k} A_k + \sqrt{1-\eta_k} E_k \, ,
\end{equation}
where $\{ E_k, E_k^\dag \}$ are effective canonical field operators
associated to the environment and memory. Equation (\ref{linear_3}) describes the
bosonic memory channel as the direct product of uncorrelated attenuating
channels, characterized by the transmissivities $\{ \eta_k \}$, which are
in turn obtained as the singular value of the matrix $f$.
In this way, we have shown that the memory channel is unitary equivalent
to a noisy channel with independent (although non-identical) noise.
This mapping has been used in \cite{LGM} (see also \cite{trans}) to compute
the capacities of the memory channel.

From an operational point of view, if the optimal
transformations are physically implemented, standard error correcting
codes, designed for uncorrelated noise, can be applied with high effectiveness.
From this point of view, the above unitaries are not only
mathematical tools for computing the channel capacities, but they are
actual pre-processing and post-processing of information that can be used
to improve the performance of standard error correcting codes.
Once the unitaries have to be physically implemented, one can pose the
question of quantifying the resources needed for their implementation.
First of all, one can notice that the above unitaries can be
constructed by combining elementary gates coupling at most two bosonic modes
(such as beam-splitters and phase-shifters).
According to \cite{Reck}, a unitary over $n$ channel
uses require at most $n(n+1)/2$ elementary transformations.
However, the decomposition provided by \cite{Reck} requires the action
of beam-splitter transformations coupling arbitrary pairs of channel outputs.
It is hence clear that to apply these transformations, one first has to wait
for $n$ channel uses and then start the process.
For example, for the perfect memory channel, $\epsilon = 1$, the memory channel is
unitary itself, and it can be perfectly inverted by waiting the $n$-th channel
output and then reverse its action.

In the following we consider a more efficient though not perfect procedure.
Namely, we consider alternative unitary operators constructed by composition of elementary
gates in a number scaling linearly with the number of channel uses.
For any $n$, these pre-processing and post-processing unitaries only couple a fixed
number $\ell$ of consecutive input and output modes.


\section{Partial unraveling of correlations}\label{partial}

We consider a scheme in which, upon $n$ uses of the channel, pre-processing
and post-processing unitaries are prepended and appended to the memory channel.
We consider the case of canonical unitaries, hence the input modes are transformed
as follows
\begin{align}
A_k & := \sum_j {\mathcal{U}^{(\ell)}}_{kj} \, a_j \, , \label{input2} \\
B_k & := \sum_j {\mathcal{V}^{(\ell)}}_{kj} \, b_j \, . \label{output2}
\end{align}

\begin{figure}[htb]
\centering
\includegraphics[width=0.48\textwidth]{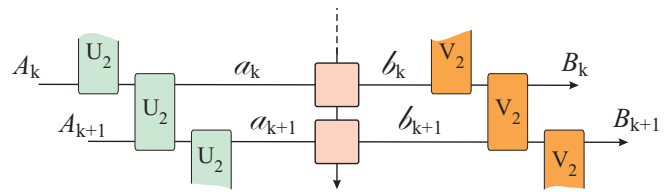}
\caption{Pre-processing and post-processing of depth $\ell=2$,
involving unitary transformations which couple two consecutive bosonic modes
at the input of the memory channel (pre-processing), and
two consecutive output modes (post-processing).}\label{depth2}
\end{figure}

\begin{figure}[htb]
\centering
\includegraphics[width=0.48\textwidth]{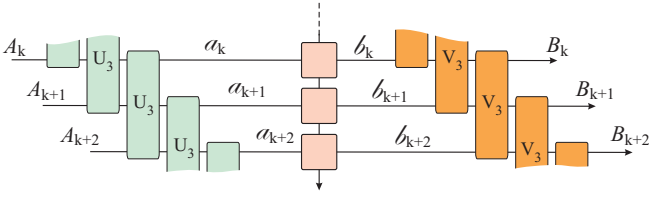}
\caption{Pre-processing and post-processing of depth $\ell=3$,
involving unitary transformations which couple three consecutive
bosonic modes at the input of the memory channel (pre-processing), and
three consecutive output modes (post-processing).
The generalization to unitaries of higher depth is straightforward.}\label{depth3}
\end{figure}

In order to consider unitaries which can be applied in a efficient way, 
we assume that the pre-processing and post-processing unitaries are obtained 
as the concatenation of elementary unitary transformations $U_\ell$, $V_\ell$, 
which only couple $\ell$ consecutive modes, in such a way that
$\ell-1$ of the output modes of each elementary transformation enter the next one. 
This is depicted in Figs.\ \ref{depth2}, \ref{depth3} for the case of unitaries of $\ell=2$ and $\ell=3$.
We refer to the integer $\ell$ as the {\it depth} of the unitary transformations.
Denoting $I_m$ the identity transformation acting on $m$ consecutive modes, we then 
have $\mathcal{U}^{(\ell)} = (I_{n-\ell} \otimes U_\ell) \cdots (I_1 \otimes U_\ell \otimes I_{n-\ell-1}) ( U_\ell \otimes I_{n-\ell})$ 
and $\mathcal{V}^{(\ell)} = (I_{n-\ell} \otimes V_\ell) \cdots (I_1 \otimes V_\ell \otimes I_{n-\ell-1} ) (V_\ell \otimes I_{n-\ell})$.

The application of the pre-processing and post-processing unitaries lead 
to the input-output relations
\begin{equation}\label{linear_4}
B_k = \sum_{j} \tilde f_{kj} \, A_j + \sum_j \tilde g_{kj} \, e_j + \tilde t_k m_1 \, ,
\end{equation}
where $\tilde f_{kj} = \sum_{l,m} {\mathcal{V}^{(\ell)}}_{kl} f_{lm} {\mathcal{U}^{(\ell)\dag}}_{mj}$,
$\tilde g_{kj} = \sum_{l,j} {\mathcal{V}^{(\ell)}}_{kl} g_{lj}$ and $\tilde t_k = \sum_l {\mathcal{V}^{(\ell)}}_{kl} t_l$.

Then, we can optimize the choice of the unitaries of a given depth, in terms of a
suitable performance quantifier. Here we would like to quantify the amount of correlations
between the $k$-th output mode and the $k$-th input mode, and between the $k$-th output mode
and all the other input modes. In the case in which $\mathcal{U}^{(\ell)} = \mathcal{U}$,
$\mathcal{V}^{(\ell)} = \mathcal{U}$, we have the unraveling of correlations, that is, the
correlations between the $k$-th output and $k$-th input are maximal, and the $k$-th output is
uncorrelated with the other inputs. Clearly, by increasing the depth of the unitaries,
one gets higher and higher performances, since the case of unitaries with depth equal
to $\ell$ is obtained as a special case of the unitaries with depth equal to $\ell' > \ell$.

We are going to use the mutual information as a quantifier of correlations.
To define the mutual information, we need to consider a specific instance of
encoding/decoding procedure for classical communication via the memory channel.
From Ref.\ \cite{vit} we know that for the corresponding memoryless attenuating channel
a standard encoding/decoding to get close to the classical capacity is given by (single mode)
coherent states and dyne measurements \cite{dyne}.

Then, we assume the input states at the $k$-th channel-use to be coherent states, with Gaussian
distributed amplitude. Explicitly, the input states at the $k$-th use of the channel are coherent states
$|\alpha_k\rangle$, whose amplitude is chosen according to the Gaussian probability density distribution
\begin{equation}
P(|\alpha_k\rangle) \simeq \exp{\left( - \frac{|\alpha_k|^2}{N}\right)}.
\end{equation}
(Here and in what follows we neglect to explicitly write the proper
normalization factor.)
The amplitudes $\alpha_k$, $\alpha_h$ are taken to be mutually
independent for $h \neq k$.
We assume the initial state of the memory mode and of the
environmental modes to be the vacuum state. Under these assumptions,
the $k$-th output of the memory channel is also a coherent state, from  Eq.\ (\ref{linear_4}),
\begin{equation}
|\beta_k\rangle = | \sum_l \tilde f_{kl} \alpha_l\rangle \, .
\end{equation}

For extracting classical information from the quantum states, we
consider the heterodyne measurement \cite{dyne} described by the POVM elements
\begin{equation}
E_{\gamma_k} = \frac{|\gamma_k\rangle\langle\gamma_k|}{\pi}.
\end{equation}
Since, in the chosen encoding scheme, the $k$-th output is
$|\beta_k\rangle$, the probability of measuring the amplitude $\gamma_k$ is
\begin{align}
P(\gamma_k|\beta_k) & = \mathrm{Tr}(E_{\gamma_k} |\beta_k\rangle\langle\beta_k|) \nonumber \\
& = \frac{|\langle\gamma_k|\beta_k\rangle|^2}{\pi} 
\simeq \exp{\left(-|\gamma_k - \beta_k|^2\right)} \, . \label{POVM}
\end{align}

We can now compute the probability density that, given the coherent
state $|\alpha_k\rangle$ is sent at the $k$-th channel use, the
amplitude $\gamma_k$ is measured at the output. That conditional
probability density is obtained by integrating the Eq.\ (\ref{POVM})
over all the $\alpha_l$ with $l \neq k$, i.e.,
\begin{equation}\label{cond}
P(\gamma_k|\alpha_k) \simeq \int
\exp{\left(-|\gamma_k-\beta_k|^2\right)} \prod_{l \neq k} 
P(|\alpha_l\rangle) d^2\alpha_l \, .
\end{equation}
We can also compute the joint probability distribution
\begin{align}
P(\gamma_k,\alpha_k) & = P(\gamma_k|\alpha_k) 
P(|\alpha_k\rangle) \nonumber \\
& = \int \exp{\left(-|\gamma_k-\beta_k|^2\right)} \prod_{l} 
P(|\alpha_l\rangle) d^2\alpha_l \, . \label{joint}
\end{align}
Using the explicit expression of the amplitude, $\beta_k = \sum_l \tilde f_{kl} \alpha_l$,
it is straightforward to compute the Gaussian integrals in Eqs.\
(\ref{cond}), (\ref{joint}).

From the probability distribution $P(\gamma_k,\alpha_k)$, it is easy
to calculate the mutual information of variables $\gamma_k$ and $\alpha_k$. 
The mutual information is (with a bit of abuse of notation)
\begin{equation}
I_k = H[P(\gamma_k)] + H[P(\alpha_k)] - H[P(\gamma_k,\alpha_k)],
\end{equation}
where $H$ denotes the Shannon entropy, and
\begin{equation}
P(\gamma_k) = \int P(\gamma_k,\alpha_k) 
P(|\alpha_k\rangle) d^2\alpha_k \, .
\end{equation}

Analogously, one can compute the mutual information between the
measured amplitude at $k$-th output and the amplitude of the coherent
states $\alpha_h$ with $h \neq k$. In order to do that, let us
denote {$\vec\alpha_{\not k}$ the complex vector whose
entries are the amplitudes $\{ \alpha_l \}_{l \neq k}$.
Hence, the conditional probability that the output measured
amplitude has value $\alpha$ is
\begin{equation}
P(\gamma_k | \vec\alpha_{\not k}) \simeq \int
\exp{\left(-|\gamma_k - \beta_k|^2\right)} 
P(|\alpha_k\rangle) d^2\alpha_k \, ,
\end{equation}
from which
\begin{equation}
P(\gamma_k , \vec\alpha_{\not k}) = P(\gamma_k |
\vec\alpha_{\not k}) \prod_{l \neq k}
P(|\alpha_l\rangle) \, .
\end{equation}
These Gaussian probability density distributions can explicitly
written in terms of the matrix coefficients $\tilde
f_{kl}$ in Eq.\ (\ref{linear_4}). The mutual information between the
measured amplitude $\gamma_k$ at the $k$-th channel output and the all
the input amplitudes $\alpha_l$, with $l \neq k$,  is
\begin{equation}
I'_{k} = H[P(\gamma_k)] + H[P(\vec\alpha_{\not k})]
- H[P(\gamma_k,\vec\alpha_{\not k})] \, .
\end{equation}

Explicitly, we have
\begin{align}
I_k & = \frac{1}{2} \left[ \log_2{M_{11}} + \log_2{M_{22}} - \log_2{\det \left( \begin{array}{cc} M_{11} & M_{12}\\ M_{21} & M_{22} \end{array}\right)} \right] \, , \\
I'_k & = \frac{1}{2} \left[ \log_2{M_{11}} + \log_2{M_{33}} - \log_2{\det \left( \begin{array}{cc} M_{11} & M_{13}\\ M_{31} & M_{33} \end{array}\right)} \right] \, ,
\end{align}
where $M_{ij}$ are the elements of the matrix
\begin{eqnarray}
M = \left(\begin{array}{ccc}
1+N(\tilde f_{kk}^2+\mu^2) & N\tilde f_{kk} & N\mu \\
N\tilde f_{kk}             & N              & 0  \\
N\mu                       & 0              & N
\end{array}\right) \, ,
\end{eqnarray}
with
\begin{equation}
\mu = \sqrt{ \sum_{j\neq k} | \tilde f_{kj} |^2 } \, .
\end{equation}

By increasing the value of $k$, the mutual informations quickly converge to the limiting functions
\begin{eqnarray}
I & := & \lim_{k\to \infty} I_k \, , \\
I' & := & \lim_{k\to \infty} I'_{k} \, .
\end{eqnarray}
Our aim is to find optimal pre-processing and post-processing unitaries, for any given $\ell$,
that can maximize $I$ and minimize $I'$.


\begin{figure}[htb]
\centering
\includegraphics[width=0.4\textwidth]{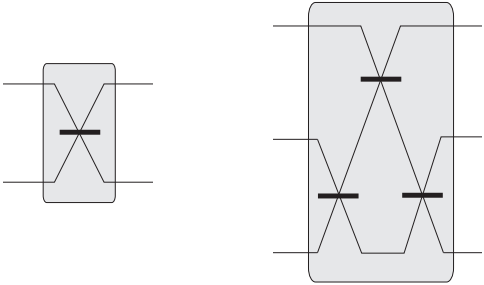}
\caption{Left: a depth-$2$ canonical unitary realized as beam-splitter transformation.
Right: a depth-$3$ canonical unitary realized as a network of three beam-splitters,
according to Euler decomposition.}\label{param}
\end{figure}


\section{Results}\label{Res}

As examples, we present results for the optmization in the case of unitaries of depth $\ell=2$ and $\ell=3$.
For the case of depth-$2$ unitaries, we assume $\mathcal{U}^{(2)}$ and  $\mathcal{V}^{(2)}$ of the form
\begin{eqnarray}
\left(\begin{array}{cc} \cos{\theta} & \sin{\theta} \\
-\sin{\theta} & \cos{\theta}
\end{array}\right)
\end{eqnarray}
with $\theta_{\mathcal U}$ and $\theta_{\mathcal V}$ respectively.

The optimization is then performed over the two angles $\theta_{\mathcal U}$, $\theta_{\mathcal V}$,
which are the parameters of the beam-splitter trasformations underlying the canonical
unitaries, see Fig.\ \ref{param}.

For the case of depth-$3$ unitaries, we consider canonical unitaries $\mathcal{U}^{(3)}$ and  $\mathcal{V}^{(3)}$
acting on three modes, parameterized according to Euler decomposition,
\begin{align}
& \left(\begin{array}{ccc}
1 & 0 & 0 \\
0 & \cos{\gamma} & \sin{\gamma} \\
0 & -\sin{\gamma} & \cos{\gamma} \end{array}\right)
\left(\begin{array}{ccc}
\cos{\theta} & \sin{\theta} & 0 \\
-\sin{\theta} & \cos{\theta} & 0 \\
0 & 0 & 1 \end{array}\right) \nonumber\\
& \times \left(\begin{array}{ccc}
1 & 0 & 0 \\
0 & \cos{\phi} & \sin{\phi} \\
0 & -\sin{\phi} & \cos{\phi} \end{array}\right)
\end{align}
with $(\gamma,\theta,\phi)_{\mathcal U}$ and $(\gamma,\theta,\phi)_{\mathcal V}$ respectively.

These transformations represent a small interferomenter composed by
a network of three beam-splitters, as depicted in Fig.\ \ref{param}.
The mutual informations are then optimized over the six angles
 $(\gamma,\theta,\phi)_{\mathcal U}$ and $(\gamma,\theta,\phi)_{\mathcal V}$.

For a given value of $N$, the maximum of $I$ and the minimum of $I'$ over the beam-splitter parameters are
plotted in Figs.\ \ref{I_2_3}, \ref{Ip_2_3}, as function of $\epsilon$ and for
different values of $\eta$. It is worth remarking that the optimal choice of the
parameters jointly maximizes $I$ and minimizes $I'$.
The figures show, as it has been remarked above, that better perfomances are
obtained for higher value of $\ell$.

Several comments are in order. First of all, the mutual information enhancement and reduction are
less pronounced for $\eta$ close to $1$: as the channel gets closer to the ideal
one, memory effects tends to disappear. Secondly, if $\eta$ is close to $0$, the
major increase and decrease in the mutual informations are for $\ell=2$, while the differences
in the mutual informations between the case with $\ell=2$ and $\ell=3$ are relatively small.
Finally, for intermediate valure of $\eta$, higher values of $\ell$ are more effective for
higher values of $\epsilon$.


\begin{figure}[htb]
\centering
\includegraphics[width=0.5\textwidth]{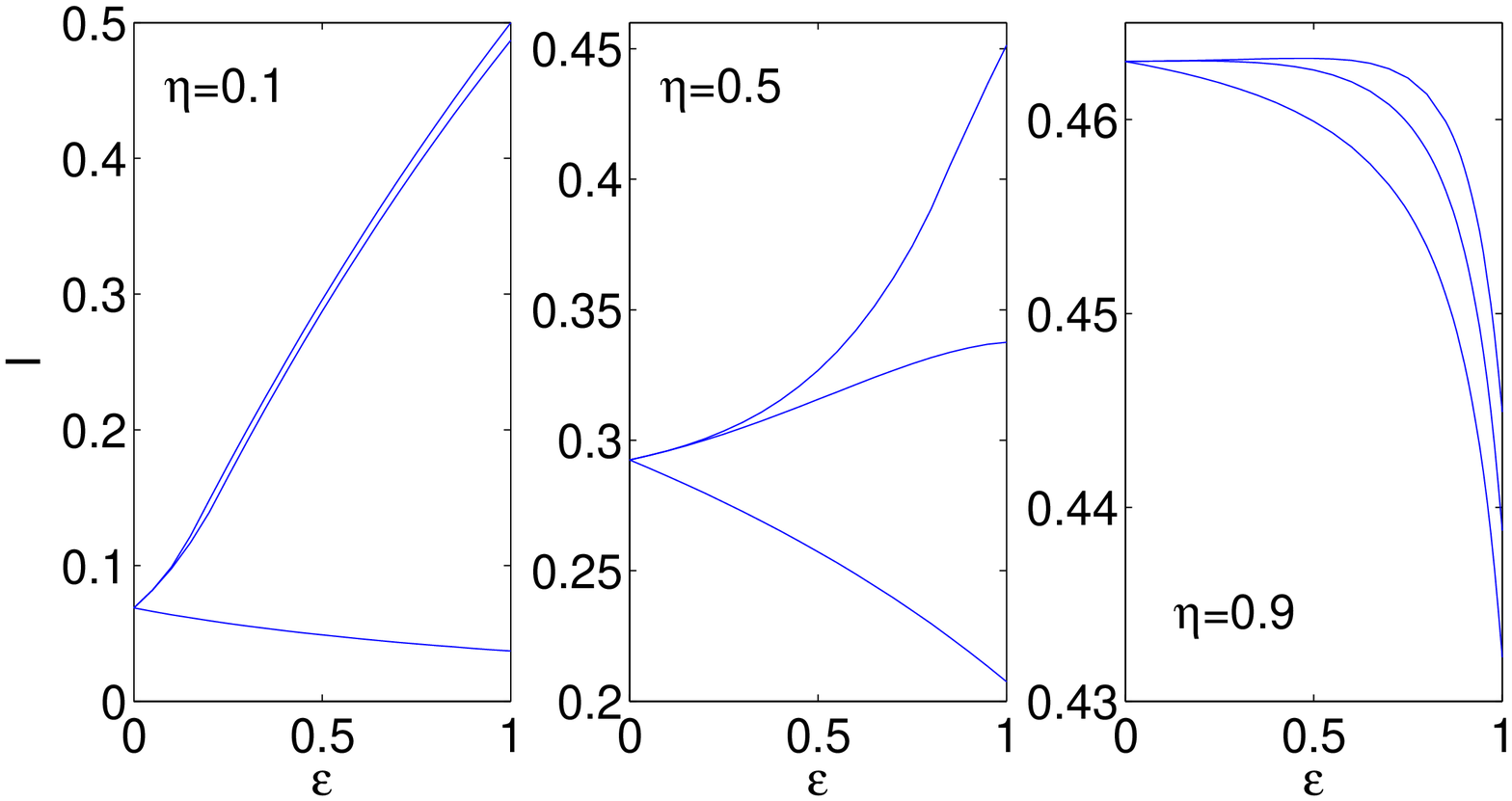}
\caption{The mutual information $I$ vs $\epsilon$, for $N=1$. From bottom to top, without
pre-processing and post-processing, with pre-processing and post-processing
of depth $\ell=2$, and with pre-processing and post-processing of depth $\ell=3$.
From left to right, the three plots refer to $\eta=0.1$, $0.5$, $0.9$.}\label{I_2_3}
\end{figure}

\begin{figure}[htb]
\centering
\includegraphics[width=0.5\textwidth]{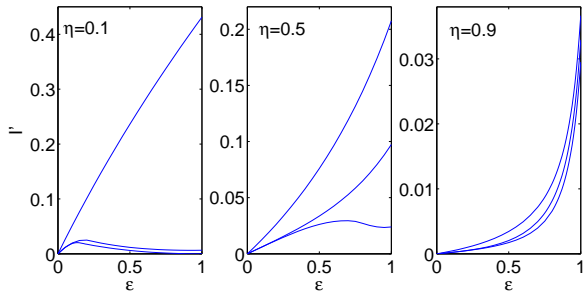}
\caption{The mutual information $I'$ vs $\epsilon$, for $N=1$. From top to bottom, without
pre-processing and post-processing, with pre-processing and post-processing
of depth $\ell=2$, and with pre-processing and post-processing of depth $\ell=3$.
From left to right, the three plots refer to $\eta=0.1$, $0.5$, $0.9$.} \label{Ip_2_3}
\end{figure}

\section{Conclusion}\label{end}

We have considered a model of quantum communication channel with memory,
i.e., characterized by non i.i.d.\ noise. We have studied the possibility of using standard
encoding/decoding strategies devised for i.i.d.\ errors by removing correlations
introduced by the quantum memory channel.
For the considered model, it is possible to identify pre-processing and post-processing
unitary transformations which allow the complete removal of correlations,
mapping the correlated noises into independent (although non identical) ones.
However, since these unitaries act globally on a train of transmitted signals,
their implementation cannot be efficient, since one has to wait the
transmission of long strings of signals before the decoding process can start.
We have hence considered some examples of unitary transformations with finite
depth, involving two or three consecutive signals, permitting the efficient
although partial reduction of the correlations introduced by the memory channel.
According to suitable a quantifier such as the mutual information,
the correlation reduction increases with increasing depth of the
unitary transformations, but already at small depths it appears significant.

By increasing the depth of the pre-processing and post-processing unitaries the
reduction of correlations gets improved, as these unitaries become closer and closer
to those allowing the complete removal of correlations.
Hence, the depth could be used to bound the error probability of standard codes.

Finally it is worth noticing that the introduced unitaries provide an online
pre-processing and post-processing procedure that combined with standard codes
resembles the convolutional  codes \cite{convECC}.
A deeper investigation of this parallelism is left for future work.


\section{acknowledgments}
We warmly thank Reinhard F. Werner, David Gross, Johannes G\"utschow, 
and Vittorio Giovannetti for useful discussions.
The research leading to these results has received funding
from the European Commission's seventh Framework
Programme (FP7/2007-2013) under Grant Agreement No.\ 213681.



\end{document}